\documentclass[natbib]{article}

\usepackage{mathtools}
\usepackage{graphicx}

\DeclareGraphicsExtensions{.pdf,.png,.jpg}

\usepackage{times}
\usepackage{graphicx} 
\usepackage{subfigure} 
\usepackage{amsmath}
\usepackage{mathtools}
\usepackage{natbib}
\usepackage{epstopdf}
\usepackage{algorithm}
\usepackage{algorithmic}

\usepackage{multirow}

\usepackage[dvipdfmx]{hyperref}

\usepackage{hyperref}

\def \tabref #1{~Table \ref{#1}~}
\def \figref #1{~Figure \ref{#1}~}

\usepackage[accepted]{icml2014}
\usepackage{listings,upquote}

\icmltitlerunning{ MILJS : Brand New JavaScript Libraries for Matrix Calculation and Machine Learning }

\begin{document} 

\twocolumn[
\icmltitle{ Implementation of a Practical Distributed Calculation System with Browsers and JavaScript, and Application to Distributed Deep Learning }

\icmlauthor{Ken Miura}{miura@mi.t.u-tokyo.ac.jp}
\icmlauthor{Tatsuya Harada}{harada@mi.t.u-tokyo.ac.jp}
\icmladdress{ Machine Intelligence Laboratory, Department of Mechano-Informatics, The University of Tokyo }

\icmlkeywords{javascript libraries, machine learning, matrix calculation, browsers, open source}
\vskip 0.3in
]

\begin{abstract}
Deep learning can achieve outstanding results in various fields. However, it requires so significant computational power that graphics processing units (GPUs) and/or numerous computers are often required for the practical application. We have developed a new distributed calculation framework called "Sashimi" that allows any computer to be used as a distribution node only by accessing a website. We have also developed a new JavaScript neural network framework called "Sukiyaki" that uses general purpose GPUs with web browsers. Sukiyaki performs 30 times faster than a conventional JavaScript library for deep convolutional neural networks (deep CNNs) learning. The combination of Sashimi and Sukiyaki, as well as new distribution algorithms, demonstrates the distributed deep learning of deep CNNs only with web browsers on various devices. The libraries that comprise the proposed methods are available under MIT license at {\em http://mil-tokyo.github.io/}.
\end{abstract}

\section{Introduction}

Utilization of big data has recently come to play an increasingly important role in various business fields. Big data is often handled with deep learning algorithms that can achieve outstanding results in various fields. For example, almost all of the teams that participated in the ILSVRC 2014 image recognition competition \cite{ILSVRC2014} used deep learning algorithms. Such algorithms are also used for speech recognition \cite{Dahl:2012,Hinton:2012} and molecular activity prediction \cite{Kaggle}.

However, deep learning algorithms have huge computational complexity and often require the use of graphics processing units (GPUs) and numerous computers for practical distributed computation. It is difficult to construct a distributed computation environment, and it frequently requires the preparation of certain operating systems and installation of specific software, e.g., Hadoop \cite{Shvachko:2010}. For practical application of deep learning algorithms, a new instant distributed calculation environment is eagerly anticipated.

\section{Sashimi : Distributed Calculation Framework}

We have developed a new distributed calculation framework called "Sashimi." In a general distributed calculation framework, it is difficult to increase the number of node computers because users must install client software on each node computer. With Sashimi, any computer can become a node computer only by accessing a certain website via a web browser without installing any client software. The proposed system can execute any code written in JavaScript in a distributed manner.

\subsection{Implementation}

\begin{figure}[!ht]
  \begin{center}
    \includegraphics[width=\columnwidth]{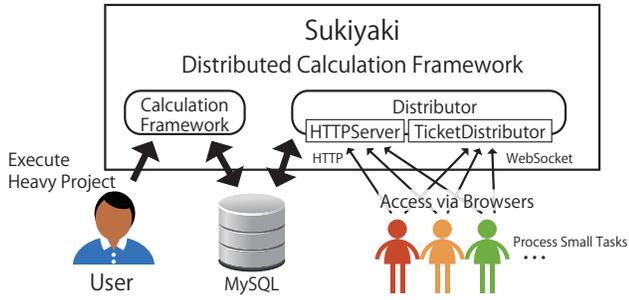}
    \caption{Sukiyaki Architecture \label{fig:architecture}}
  \end{center}
\end{figure}

Sashimi consists of two servers, the CalculationFramework and Distributor servers (\figref{fig:architecture}). When a user runs a project that includes distributed processing using the CalculationFramework and accesses the Distributor via web browsers, the processes can be distributed and executed in multiple web browsers. The results processed by the distributed machines can be used as if they were processed by a local machine without being conscious of their differences by using the CalculationFramework.

\vspace{-0.5mm}
\vspace{-0.5mm}

\subsubsection{CalculationFramework}

CalculationFramework describes calculations that include distributed processing. If a user writes code according to a certain interface, the processes are distributed and computed in multiple web browsers via Distributor. The results computed in the web browsers are then automatically collected. This allows the results to be used as if they were processed by the local machine. In this subsection, we introduce PrimeListMakerProject, which finds prime numbers from 1 to 10,000, as an example.

A "project" is a programming unit of CalculationFramework that has an endpoint from which a process starts. In a project, a user can execute distributed processing by creating a task instance. Note that processes that do not require distributed processing are also supported.

A "task" is a process that is distributed and executed in web browsers. If a user writes a task according to a certain interface, the arguments are automatically divided and distributed to web browsers. The processed results are then automatically collected. In PrimeListMakerProject, the task that determines whether an input integer is a prime number is called IsPrimeTask. This task is distributed among web browsers. The task is given arguments generated by the project, and it must return the calculated results using a callback function. Note that the user can use external libraries and datasets. In this example, the task calls is\_prime function in an external library, which determines whether the input integer is a prime number.

After the project generates the task instance with arguments, the framework generates tickets for each divided argument. The framework sends the codes and the arguments as tickets with the external libraries and datasets to the Distributor via MySQL. The tickets distributed by the Distributor are collected and can be used again by the CalculationFramework via MySQL. Since the project is implemented with server-side JavaScript Node.js and the task is implemented with JavaScript, they can be used without considering whether the code is executed by the server or the browsers.

\subsubsection{Distributor}

The Distributor distributes tasks and tickets, which are sent from the CalculationFramework via MySQL, to browsers. The Distributor also collects the results calculated in the browsers. Note that the Distributor consists of two servers, i.e., HTTPServer and TicketDistributor.

HTTPServer, which is a web server implemented with Node.js, provides static files that include a basic program and discloses APIs that offer datasets to be used in the distributed calculation. If a user wants to make a browser function as a node, the user only needs to access the basic program provided by HTTPServer via the web browser. The basic program consists of a static HTML file and a JavaScript file. The basic program works as follows.

\begin{enumerate}
\item a connection with TicketDistributor is established using WebSocket
\item a ticket request is sent to TicketDistributor
\item a task request is sent to TicketDistributor if it has not downloaded the task described in the ticket
\item a request for required external datasets and files is sent to HTTPServer
\item the task with arguments described in the ticket is executed
\item the calculated result is returned to TicketDistributor
\item return to Step 2.
\end{enumerate}

The task and external data are cached in the browser. If a program runs for a long time, memory usage increases due to the cache. Therefore, we have implemented garbage collection on the basis of the least recently used algorithm. If an error occurs when the task is running, an error report that includes a stack trace is sent to the TicketDistributor. Then, the browser reloads itself. Thus, a task described by the tickets generated by the CalculationFramework can be continuously executed without special maintenance once the user accesses the program.

Users can check the progress of a task and tickets via the HTTPServer control console. In this console, users can see the project name, the number of tasks, the number of tickets waiting to be processed, the number of executed tickets, the number of error reports, and the client information for the project. Note that a console that can be used to execute code in web browsers is also provided. With this console, the user can make the browsers reload themselves and redirect to another distributed system.

We use responsive web design (RWD) techniques in the user interface (UI) of the basic program and the control console. RWD techniques adapt the UI to the screen size of any PC, tablet, or smartphone, which makes it easy to use any devices for distributed calculation and check of the progress.

The tasks and tickets generated by the CalculationFramework are distributed to browsers via WebSocket by the TicketDistributor. The processed results are also collected by the TicketDistributor. Unlike HTTPServer, the TicketDistributor runs in a single process and communicates with each web browser unitarily and efficiently.

When the TicketDistributor receives a ticket request from the browser, it obtains tickets in ascending order of "virtual created time" from the MySQL server. Virtual created time is determined as follows.

\begin{itemize}
\item the virtual created time is the ticket creation time of undistributed tickets
\item virtual created time is five minutes after ticket distribution if the tickets have been distributed
\item virtual created time is five minutes after the last ticket distribution if the ticket has been redistributed
\end{itemize}

Thus, if the results are not returned within five minutes, the tickets is treated in such a way as to be re-created. Note that tickets are redistributed in ascending order by distribution time if there are no further tickets to be distributed. Thus, if a web browser is terminated after it receives a ticket, and/or if there are clients with low computational capability, another client can execute the task. Therefore, the throughput can be enhanced. The tickets are redistributed at intervals of at least 10 seconds, which prevents the last ticket from being distributed to many clients and prevents the next calculation from being delayed. We implemented this algorithm using SQL, which can quickly select tickets to be distributed.

\subsection{Benchmark}
\subsubsection{Experiment Condition}

Using Sashimi, we demonstrate that a task with high computational cost can be computed in parallel efficiently. Here, we compare the time required to classify the MNIST dataset with a nearest neighbor method by changing the number of clients. In this experiment, 1,000 images from the 10,000 MNIST test images were classified by comparing them with 60,000 training images. We used one to four clients on a desktop computer and tablet PCs described in \tabref{tab:mnistbench}. We accessed the Distributor using the Google Chrome web browser on both the desktop and tablet environments.

\begin{table}[!t]
	\centering
	\caption{Specifications of Devices for Distributed MNIST Benchmark}
	\vspace {0.3mm}
	\begin{tabular}{| r | c | c |} \hline
					& DELL OPTIPLEX			& Nexus 7			\\ \hline \hline
		Model		& DELL OPTIPLEX 8010		& Nexus 7 (2013)	\\ \hline
		OS			& Windows 7 Professional		& Android 4.4.4		\\ \hline
		CPU			& Intel Core i7-3770 3.4GHz	& Krait 1.50GHz	\\ \hline
		RAM			& 16GB					& 2GB			\\ \hline
	\end{tabular}
	\label{tab:mnistbench}
\end{table}

\subsubsection{Results}

The results are shown in \tabref{tab:mnistbench_result}. In both environments, the calculation time was reduced with the distributed computation. The effect of the distributed computation was remarkable when the proposed system was used with the tablet PC because the tablet has lower computational power than the desktop computer and the overhead time required for the distribution becomes relatively shorter. We believe that the proposed distributed computing method will become more effective for other feature extraction methods with high computational costs such as SIFT and deep learning.

\begin{table*}[!t]
	\centering
	\caption{Results of Distributed MNIST Benchmark}
	\vspace {0.3mm}
	\begin{tabular}{| c | r | c | c |} \hline
		Environment	& Clients	& Elapsed Time (sec.)	& Elapsed Time (ratio)	\\ \hline \hline
		\multirow{4}{*}{DELL OPTIPLEX}
					& 1		& 107				& 1					\\ \cline{2-4}
					& 2		& 62					& 0.58				\\ \cline{2-4}
					& 3		& 52					& 0.49				\\ \cline{2-4}
					& 4		& 46					& 0.43				\\ \hline
		\multirow{4}{*}{Nexus 7}
					& 1		& 768				& 1					\\ \cline{2-4}
					& 2		& 413				& 0.54				\\ \cline{2-4}
					& 3		& 293				& 0.38				\\ \cline{2-4}
					& 4		& 255				& 0.33				\\ \hline
	\end{tabular}
	\label{tab:mnistbench_result}
\end{table*}

\section{Sukiyaki : Deep Neural Network Framework}

We implemented a learning algorithm for deep neural networks (DNNs) with browsers based on Sashimi. Here, we explain the proposed framework and implementation for DNNs. We also discuss the advantages of the proposed method over an existing library in a stand-alone environment. The distributed computation of DNNs is explained in the next section.

We primarily implemented deep convolutional neural networks (deep CNNs) that obtain high classification accuracy in image recognition tasks. ConvNetJS \cite{ConvNetJS} was implemented as a NN library using JavaScript. However, its computational speed is limited because it runs on only a single thread. Therefore, we developed a deep neural network framework called "Sukiyaki" that utilizes a fast matrix library called Sushi \cite{Miura:2015}. The Sushi matrix library is fast because it is implemented on WebCL and can utilize general purpose GPUs (GPGPUs) efficiently.

\subsection{Implementation}

The Sukiyaki DNN framework consists of the Sukiyaki object, which handles procedures for learning and testing in the neural network, and layer objects. In this version, for deep CNNs, we implemented a convolutional layer, a max pooling layer, a fully-connected layer, and an activation layer. Note that we can add other layers if we implement certain methods such as forward, backward, and update. The forward, backward, and update methods in each layer are implemented using the Sushi matrix library; thus, they can be executed in parallel on GPGPUs.

We can use AdaGrad \cite{Duchi:2011} as an online parameter leaning method, which can learn parameters quickly. The original update rule in AdaGrad is as follows.
\begin{eqnarray*}
\theta_{i, t} = \theta_{i, t-1} - \frac{\alpha}{\sqrt{\sum_{u=0}^{t} g_{i, u}^2}} g_{i, t}
\end{eqnarray*}
where, $ \alpha $ is a scalar learning rate, $ \theta_{i, t} $ is the i-th element of the parameter at time step $t$ and $ g_{i, t} $ is the i-th element of the gradient at time step $t$. However, in this update rule, learning usually becomes unstable because the sum of squared gradients is minuscule early in the learning process. Therefore, we have modified the update rule using a constant $ \beta $ as follows.
\begin{eqnarray*}
\theta_{i, t} = \theta_{i, t-1} - \frac{\alpha}{\sqrt{\beta + \sum_{u=0}^{t} g_{i, u}^2}} g_{i, t}
\end{eqnarray*}

We designed the Sukiyaki DNN framework to be used with both Node.js and browsers so DNNs can be trained in a distributed manner using the Sashimi distributed calculation framework. For example, a model file wherein the parameters are encoded with base64 is formatted in JSON. Note that although the model file is a platform independent string format, it can be exchanged among machines without rounding errors.

\subsection{Benchmark}
\subsubsection{Experiment Condition}

We compared the learning speed of the Sukiyaki DNN framework with that of ConvNetJS, which is an existing JavaScript NN library.

\begin{figure*}[!ht]
  \begin{center}
    \includegraphics[width=14cm]{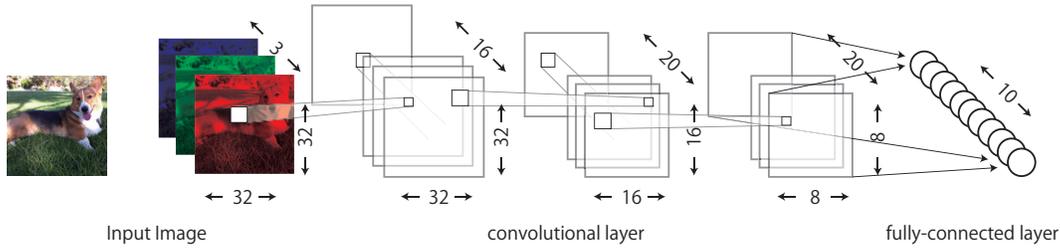}
    \caption{Deep CNN for the Benchmark of Stand-alone Deep Learning \label{fig:standalone_network}}
  \end{center}
\end{figure*}

In this experiment, we used the deep CNN model shown in \figref{fig:standalone_network}. Fifty images per mini-batch were learned from the 50,000 training images in cifar-10 \cite{Krizhevsky:2009}. Note that cifar-10 consists of 24-bit 32 $\times$ 32 color images in 10 classes. The model convolves the input images with 5 $ \times $ 5 kernels in each convolutional layer and generates three feature maps of size $32 \times 32 \times 16$, $16 \times 16 \times 20$, and $8 \times 8 \times 20$. Each convolutional layer is followed by an activation layer and a max pooling layer, and the size of the output is halved. The fourth layer is a fully-connected layer that converts 320 input elements to 10 output elements by estimating class probabilities via the softmax function.

We trained the network using Node.js and the Firefox web browser with a MacBook Pro (specs described in \tabref{tab:specification_for_deepCNN_bench}) and compared the learning speeds.

\vspace{-0.5mm}
\vspace{-0.5mm}

\begin{table}[!t]
	\centering
	\caption{Specifications of Device for Neural Network Libraries Benchmark}
	\vspace {0.3mm}
	\begin{tabular}{| r | c | c |} \hline
		Model		& MacBook Pro (Retina, 13-inch, Late 2013)	\\ \hline
		OS			& Mac OS X Yosemite (10.10.1)			\\ \hline
		CPU			& Intel Core i5-4258U 2.4GHz				\\ \hline
		GPU			& Intel Iris								\\ \hline
		RAM			& 8GB								\\ \hline
	\end{tabular}
	\label{tab:specification_for_deepCNN_bench}
\end{table}

\subsubsection{Results}

\begin{figure}[!t]
  \begin{center}
    \includegraphics[width=\columnwidth]{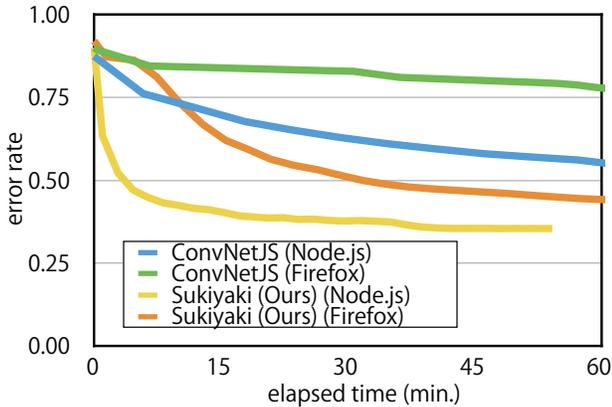}
    \caption{Error Rate \label{fig:sukiyaki_standalone_error_rate}}
  \end{center}
\end{figure}

The results are shown in \tabref{tab:sukiyaki_standalone_time_consumption} and \figref{fig:sukiyaki_standalone_error_rate}. As observed, Sukiyaki learned the network faster than ConvNetJS for both Node.js and Firefox. The convergence speed of Sukiyaki was also faster than that of ConvNetJS. Note that Sukiyaki with Node.js learned the network 30 times faster than ConvNetJS.

\begin{table}[!t]
	\centering
	\caption{Numbers of Batches Learned per 1 min.}
	\vspace {0.3mm}
	\begin{tabular}{| c | c | c | c |} \hline
		\multicolumn{2}{|c|}{ConvNetJS} 	& \multicolumn{2}{|c|}{Sukiyaki}		\\ \hline
		Node.js			& Firefox		& Node.js			& Firefox		\\ \hline \hline
		17.55			& 2.44		& 545.39			& 31.39		\\ \hline
	\end{tabular}
	\label{tab:sukiyaki_standalone_time_consumption}
\end{table}

\vspace{-0.5mm}
\vspace{-0.5mm}

\section{Distributed Deep Learning}

We ran the Sukiyaki DNN framework on the Sashimi distributed calculation framework and realized distributed learning of deep CNNs.

\subsection{Distribution Algorithms}

Some previous studies have proposed methods for distributed learning of DNNs and deep CNNs.
 
Dean et al. \cite{Dean:2012} proposed DistBelief, which is an efficient distributed computing method for DNNs. In DistBelief, a network is partitioned into some subnetworks, and different machines are responsible for computation of the different subnetworks. The nodes with edges that cross partition boundaries must share their state information between machines. However, since we consider machines connected via the Internet with slow throughput, it is difficult to share the nodes' state between different machines with the proposed framework. DistBelief focuses on a fully-connected network; thus, it is also difficult to directly apply this approach to convolutional networks that share weights among different nodes.

Meeds et al. \cite{Meeds:2014} developed MLitB, wherein different training data batches are assigned to different clients. The clients compute gradients and send them to the master that computes a weighted average of gradients from all clients and updates the network. The new network weights are sent to the clients, and the clients then restart to compute gradients on the basis of the new weights. This approach is simple and easy to implement; however, it must communicate all network weights and gradients between the master and the clients. Thus, the communication overhead becomes excessively large with a large network.

Krizhevsky \cite{Krizhevsky:2014} proposed a method to parallelize the training of deep CNNs using model parallelism and data parallelism efficiently. Generally, deep CNNs consist of many convolutional layers and a few fully-connected layers. Due to weight sharing, convolutional layers incur significant computational cost relative to the small number of parameters. However, fully-connected layers have many more parameters than convolutional layers and less computational complexity. Krizhevsky \cite{Krizhevsky:2014} developed an efficient method to parallelize the training of deep CNNs by applying data parallelism in the convolutional layers and model parallelism in the fully-connected layers. However, we focus on distributed computation via the Internet; thus, we must reduce communication costs in the proposed framework.

He et al. \cite{He:2015} implemented another effective method for distributed deep learning. They parallelize the training of convolutional layers using data parallelism on multi-GPUs. Then the GPUs are synchronized and the fully-connected layers are trained on a single GPU. The computational complexity for training fully-connected layers is relatively small so that the model parallelism of fully-connected layers does not necessarily contribute to fast learning. This method of combining parallelized and stand-alone learning works efficiently and is easy to implement. However, this method has some computational resources stay idle while fully-connected layers are learned on a single GPU and there is still room for improvement.

SINGA \cite{SINGA} is a distributed deep learning platform. It supports both model partition and data partition, and we can manage them automatically through distributed array data structure without much awareness of the array partition. While SINGA is designed to accelerate deep learning by using MPI, it is unclear whether this approach is appropriate for distributed calculation via the Internet.

In this study, we have developed a new method, which parallelizes the training of convolutional layers using data parallelism and does not apply model parallelism to fully-connected layers. The proposed method trains fully-connected layers on the server while the clients train the convolutional layers. Unlike the method of He et al. \cite{He:2015}, convolutional layers and fully-connected layers are learned concurrently. This method reduces communication cost among machines and utilizes the computational capability of the server while it awaits responses from the clients.

\subsection{Benchmark}
\subsubsection{Experiment Conditions}

\begin{figure*}[!t]
  \begin{center}
    \includegraphics[width=\textwidth]{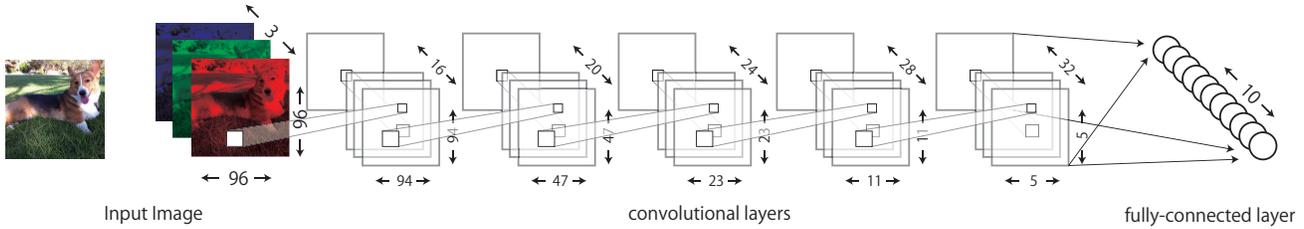}
    \caption{Deep CNN for the Distributed Deep Learning Benchmark \label{fig:distributed_network}}
  \end{center}
\end{figure*}

In this experiment, we parallelized the training of the deep CNN shown in \figref{fig:distributed_network} using the Sukiyaki DNN framework and the Sashimi distributed calculation framework. We used the computer shown in \tabref{tab:spec_distributed} as the server and clients. The client machine had four GPU cores; thus, we ran one to four Firefox web browsers on the client machine. The browsers accessed the server running Node.js and began training the network in parallel. We compared the training speeds of both convolution and fully-connected layers by varying the number of clients.

\vspace{-0.5mm}
\vspace{-0.5mm}

\begin{table*}[!t]
	\centering
	\caption{Specifications of Devices for Distributed Deep Learning}
	\vspace {0.3mm}
	\begin{tabular}{| r | c | c |} \hline
					& Server						& Client								\\ \hline
					& Mac Pro						& DELL Alienware						\\ \hline \hline
		Model		& Mac Pro (Late 2013)			& DELL Alienware Area-51				\\ \hline
		OS			& Mac OS X Yosemite (10.10.1)	& Windows 8.1							\\ \hline
		CPU			& Intel Xeon E5 3.5GHz 6-core		& Intel Core i7-5960X 3.00GHz 8-core		\\ \hline
		GPU			& AMD FirePro D500				& NVIDIA GeForce GTX TITAN Z $ \times $ 2	\\ \hline
		RAM			& 32GB						& 32GB								\\ \hline
	\end{tabular}
	\label{tab:spec_distributed}
\end{table*}

\subsubsection{Results}

\begin{figure}[!t]
  \begin{center}
    \includegraphics[width=\columnwidth]{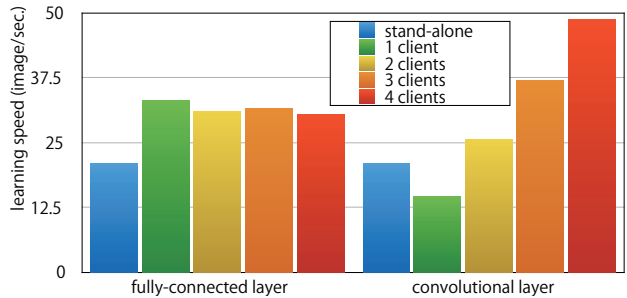}
    \caption{Learning Speed by Distributed Deep Learning \label{fig:sukiyaki_distributed_result}}
  \end{center}
\end{figure}

The results are shown in \figref{fig:sukiyaki_distributed_result}. The proposed distributed computation method can train fully-connected layers 1.5 times faster than the stand-alone computation method independently of the number of clients because the server can be devoted to training fully-connected layers. The training speed of the convolutional layers becomes faster in proportion to the number of clients. With four clients to train convolutional layers, the proposed method is two times as fast as the stand-alone method.

\vspace{-0.5mm}
\vspace{-0.5mm}

\section{Conclusion and Future Plans}

In this paper, we have proposed a distributed calculation framework using JavaScript to address the increasing need for computational resources required to process big data. We have developed the Sashimi distributed calculation framework, which can allow any web browser to function as a computation node. By applying Sashimi to an image classification problem, the task can be executed in parallel, and its calculation speed becomes faster in proportion to the number of clients. Furthermore, we developed the Sukiyaki DNN framework, which utilizes GPGPUs and can train deep CNNs 30 times faster than a conventional JavaScript DNN library. By building Sukiyaki on Sashimi, we also developed a parallel computing method for deep CNNs that is suitable for distributed computing via the Internet. We have shown that deep CNNs can be trained in parallel using only web browsers.

These libraries are available under MIT license at {\em http://mil-tokyo.github.io/}. In future, we plan to improve the efficiency of the distribution algorithm by considering clients' computational capabilities and supporting another network in Sukiyaki. Note that we welcome suggestions for improvements to our code and documentation. It is our hope that many programmers will further develop Sukiyaki and Sashimi to become a high-performance distributed computing platform that anyone can use easily.


\bibliographystyle{icml2014}
\bibliography{paper}
\onecolumn
\clearpage

\appendix

\def\thesection{Appendix}
\section{Sashimi Sample Program}

\begin{lstlisting}[basicstyle=\ttfamily\scriptsize, frame=single, caption=prime\_list\_maker\_project.js]
var ProjectBase = require('../project_base');
var IsPrimeTask = require('./is_prime_task');
var PrimeListMakerProject = function() {	
	this.name = 'PrimeListMakerProject';
	
	this.run = function(){
		var task = this.createTask(IsPrimeTask);
		var inputs = [];
		for (var i = 1; i <= 10000; i++) {
			inputs.push({ candidate : i });
		}
		task.calculate(inputs);
		task.block(function(results) {
			for (var i = 0; i < results.length; i++) {
				if (results[i].output.is_prime) {
					console.log(i + ' is a prime number.')
				}
			}
		});
	};
};
PrimeListMakerProject.prototype = new ProjectBase();
\end{lstlisting}

\begin{lstlisting}[basicstyle=\ttfamily\scriptsize, frame=single, caption=is\_prime\_task.js]
var TaskBase = require('../task_base');

var IsPrimeTask = function() {	
	this.static_code_files = ['is_prime'];

	this.run = function(input, output) {
		if (is_prime(input.candidate)) {
			output({ is_prime : true });
		} else {
			output({ is_prime : false });
		}
	};
};
IsPrimeTask.prototype = new TaskBase();
\end{lstlisting}

\begin{lstlisting}[basicstyle=\ttfamily\scriptsize, frame=single, caption=is\_prime.js]
function is_prime(candidate) {
	for (var i = 2; i <= Math.sqrt(candidate); i++) {
		if (candidate % i === 0) {
			return false;
		}
	}
	return true;
}

\end{lstlisting}

\end{document}